\documentclass[journal]{IEEEtran}
\IEEEoverridecommandlockouts
\usepackage{lipsum}
\usepackage{amsmath,amssymb,amsfonts}
\usepackage{algorithmic}
\usepackage{graphicx}
\usepackage{float}
\usepackage{subcaption}
\usepackage{cite}
\usepackage{multirow}
\usepackage{booktabs}
\usepackage{makecell}
\usepackage{textcomp}
\usepackage{xcolor}
\usepackage{url}
\usepackage{caption}
\def\BibTeX{{\rm B\kern-.05em{\sc i\kern-.025em b}\kern-.08em
    T\kern-.1667em\lower.7ex\hbox{E}\kern-.125emX}}

\begin{document}

\title{Dolphin broadband transient vocalization detection in impulsive ambient noise}

\author{Hari Vishnu$^*$, Jin Xinhui$^*$, Mandar Chitre$^{*\dagger}$, Simen Hexeberg$^*$, Koay Teong Beng$^*$
\thanks{$^*$Acoustic Research Laboratory, Tropical Marine Science Institute, National University of Singapore.}%
\thanks{$^{\dagger}$Department of Electrical and Computer Engineering, National University of Singapore.}}

\maketitle

\begin{abstract}
Passive acoustic monitoring in tropical waters such as those in Southeast Asia is hampered by the persistent, high-level impulsive noise of snapping shrimp. This is a particularly daunting challenge for detecting broadband transient vocalizations of dolphins --- clicks, buzzes and burst-pulses. These vocalizations, unlike frequency-modulated whistles, share the broadband, impulsive character of the shrimp snaps and cannot be easily separated from them by frequency-localized or energy-based means. We develop TRIDENT, a lightweight waveform-based detector for these vocalizations, and apply it to 54.2 recorder-months of data from ten island sites around Singapore. We show that the training-data pipeline and the architecture are both important: controlled signal-to-noise-ratio based mixing to generate training data, appropriate normalization and hard negative mining, together steer the detector toward learning transient temporal structure rather than  energy shortcuts. A single joint detector handles all three signal types better than specialized models for each, and outperforms established baselines. Applied to long-term recordings, it reveals tightly clustered, predominantly daytime activity, motivating days-with-detections as a robust indicator of cetacean presence.
\end{abstract}

\begin{IEEEkeywords}
Dolphin, click, burst-pulse, snapping shrimp, passive acoustics, impulsive noise, convolutional neural network, Singapore.
\end{IEEEkeywords}

\section{Introduction}
The presence of marine mammals such as dolphins indicates the health of the marine ecosystem \cite{ Alves2023}. Our study focuses on the warm coastal waters in Southeast Asia, exemplified by the hyper-urbanized waters of Singapore, which host several delphinid species --- predominantly Indo-Pacific humpback dolphins (\textit{Sousa chinensis}; IPHD) and bottlenose dolphins (\textit{Tursiops aduncus}; IPBD)~\cite{Ng2026}, but also Irrawaddy dolphins (\textit{Orcaella brevirostris}, more commonly spotted in neighbouring waters) \cite{SBR3, Kamaruzzan2013, Muhamad2022}, long-beaked common dolphins, false killer whales and finless porpoises \cite{Ming1990}. Despite these sightings, there are still no reliable estimates of the local population or visitation patterns of most of these species \cite{Ng2026, Marsh2002, Ng2022}, and the lack of observations and conclusive data  is an impediment to effective policy-making and conservation --- a problem shared across much of developing or developed Southeast Asia \cite{Marsh2019, Piwetz2021, Ponnampalam2015}. Furthermore, concern on potential decreasing pod size have been reported over the past three decades in Singapore waters, necessitating urgent and additional action to study these vital habitats of these species~\cite{Ng2026}.

IPBDs, IPHDs and Irrawaddy dolphins produce vocalizations classified as whistles, click-trains, buzzes and burst pulses \cite{Sims2012, Wang2022, Yuan2021, Caruso2020a}. 
Passive acoustic monitoring (PAM) is one of the best-available techniques for detecting and monitoring vocalizing marine mammals~\cite{Fleishman2023, Au2016, Akamatsu2001}, offering non-invasive, wide-area, around-the-clock coverage unhindered by bad visibility, time of the day, cloud cover or rough weather~\cite{Bopardikar2025}. It is a great complement to, and may even offer better detection performance than visual surveys \cite{Rayment2017} in some cases. It is particularly advantageous in the turbid waters around Singapore where visual detection is difficult -- a challenge faced in neighbouring regions as well~\cite{Bopardikar2025}. We refer the reader to prior work in \cite{Vishnu2024} and references therein for a fuller review of PAM and its usage in the Southeast Asia region. Interpreting the resulting long-term recordings from PAM systems, however, requires sifting through very large volumes of data, which makes automated, machine-learning-based detection essential \cite{Fleishman2023, Shiu2020}.

The tropical waters around Singapore pose a challenge due to their high ambient noise. Above 2~kHz the soundscape is dominated by snapping shrimp, whose colonies produce impulsive snaps with source levels upto 187~dB~re~1$\mu$Pa \cite{Kim2011, Au1998} and band levels sometimes exceeding 70~dB~re~1$\mu Pa^2$ at 10~kHz. These can drastically hinder detection of underwater signals in this region~\cite{Vishnu2024, Mahmood2017}. Snapping shrimp noise has been characterized worldwide \cite{Au1998, Ferguson2001, Readhead1997}, and is typical in Southeast Asia, but tackling its overlap with dolphin broadband transients specifically has received little attention.

In earlier work, we developed DEVMAN, a detector for the frequency-modulated (FM) vocalizations of marine mammals in Singapore's waters, with predominant focus on whistles or chirp-squeaks \cite{Vishnu2024}. Whistles concentrate their energy along narrow time-frequency trajectories, so they remain relatively separable from broadband shrimp snaps on a spectrogram. Additionally, this separability can be further enhanced through nonlinear denoising of the impulsive noise \cite{Vishnu2024, Mahmood2017}. Given the contamination from ship-based FM noise in the recordings, broadband transient vocalizations such as clicks, buzzes and burst-pulses can serve as complementary and informative indicators of dolphin presence and behaviour in addition to whistles. However, they present the opposite situation as whistles, as they share characteristics with snapping shrimp noise, namely short duration, broadband content, and impulsive structure. The target and the dominant interference therefore overlap in exactly the domain a spectrogram-and-denoising approach relies on, and the strategy that succeeded for whistles does not transfer directly. This motivates a fundamentally different, waveform-centric detection approach, which is the subject of this paper.

\begin{figure}[t]
    \includegraphics[width=\linewidth]{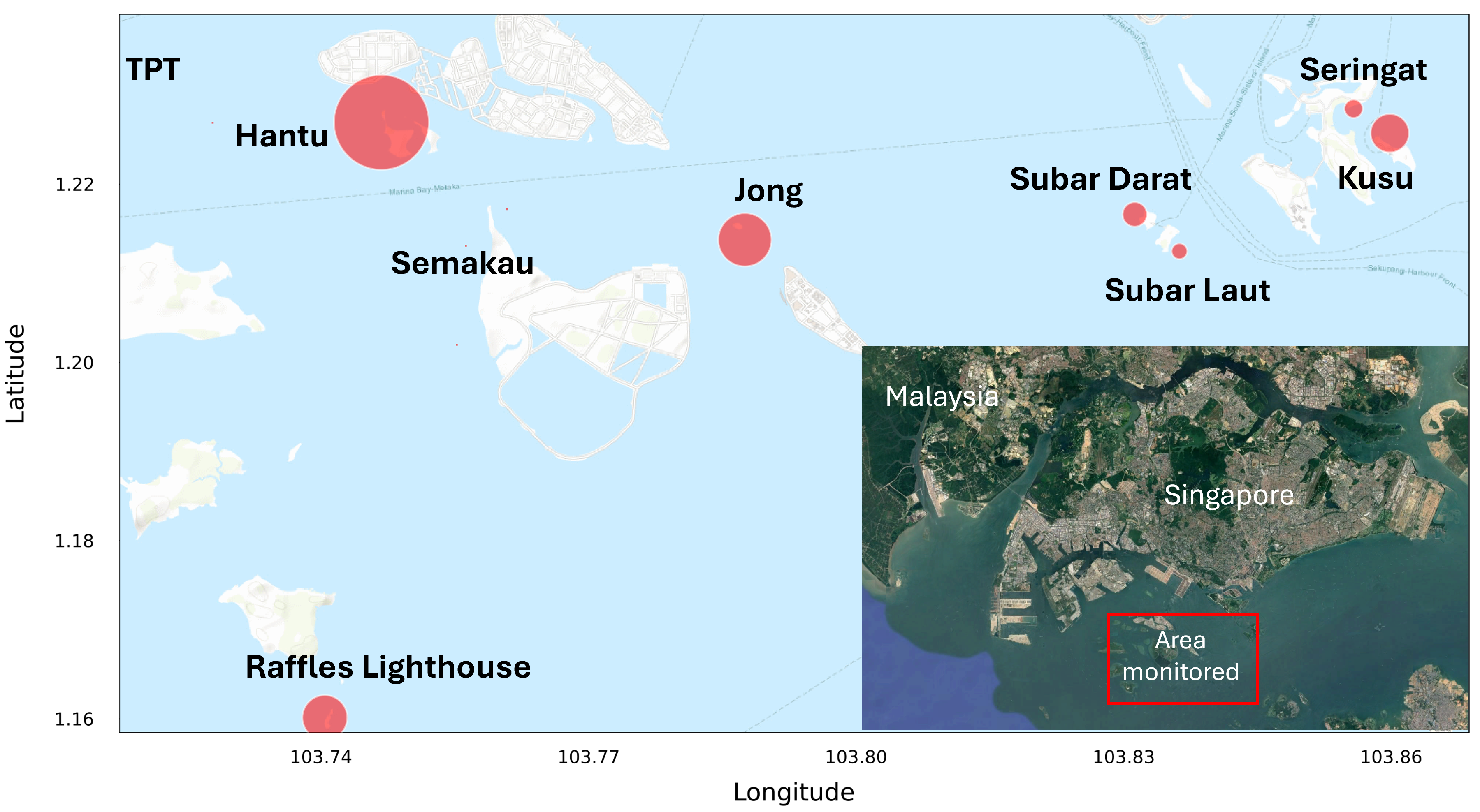}
    \caption{Map of island sites in Singapore's southern waters where passive acoustic recorders were deployed for the study, with circles whose size is proportional to number of days with detections at these sites normalized to total number of deployment days at the site, and (inset) zoomed out map of the region around Singapore showing the area studied marked by a red box.}
    \label{fig:Ls1sites}
\end{figure}

We develop TRIDENT (TRansients in Impulsive-noise DEtection NeTwork), a waveform-based one-dimensional (1D) convolutional neural network (CNN) detector for dolphin click trains, buzzes and burst-pulses, and apply it to 54.2 recorder-months of data collected across ten island sites around Singapore (shown in Fig.~\ref{fig:Ls1sites}). We show that a data-centric training strategy 
is as decisive as the architecture itself in this acoustic regime. We deliberately avoid the nonlinear denoising used for whistles to preserve the transient structure the detector relies on. Section~\ref{sec:challenges} examines the challenges of detecting broadband transients in shrimp noise and reviews existing methods, section~\ref{sec:expdesign} describes the recording setup, section~\ref{sec:detector} presents the detector design and training, section~\ref{sec:results} presents the results, and section~\ref{sec:conclusion} discusses them and concludes.

\begin{figure*}[htbp]
    \centering
    \includegraphics[width=0.9\linewidth]{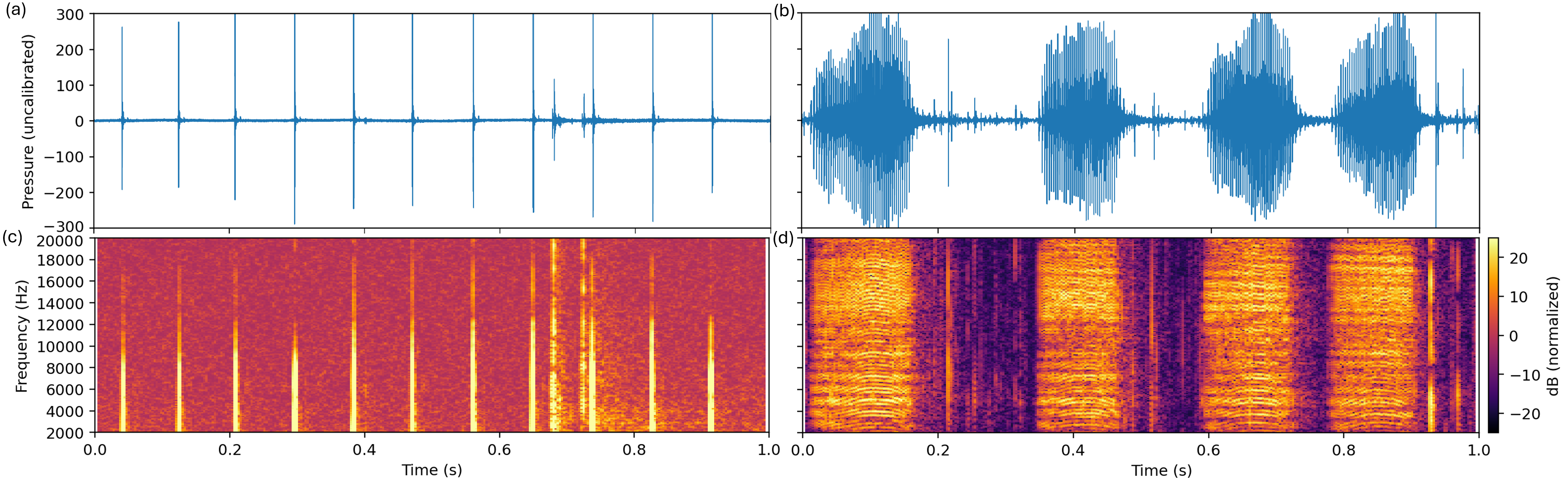}
    \caption{Representative timeseries and corresponding spectrograms of high-SNR examples of the two target signal classes considered in this work, recorded in controlled conditions. (a) and (c) Clicks are characterized by isolated broadband impulsive peaks arranged in click trains, while (b) and (d) burst-pulses exhibit more extended structure with denser temporal organization.}
    \label{fig:target_signal_examples}
\end{figure*}
\begin{figure*}[t]
    \centering
    \includegraphics[width=5in]{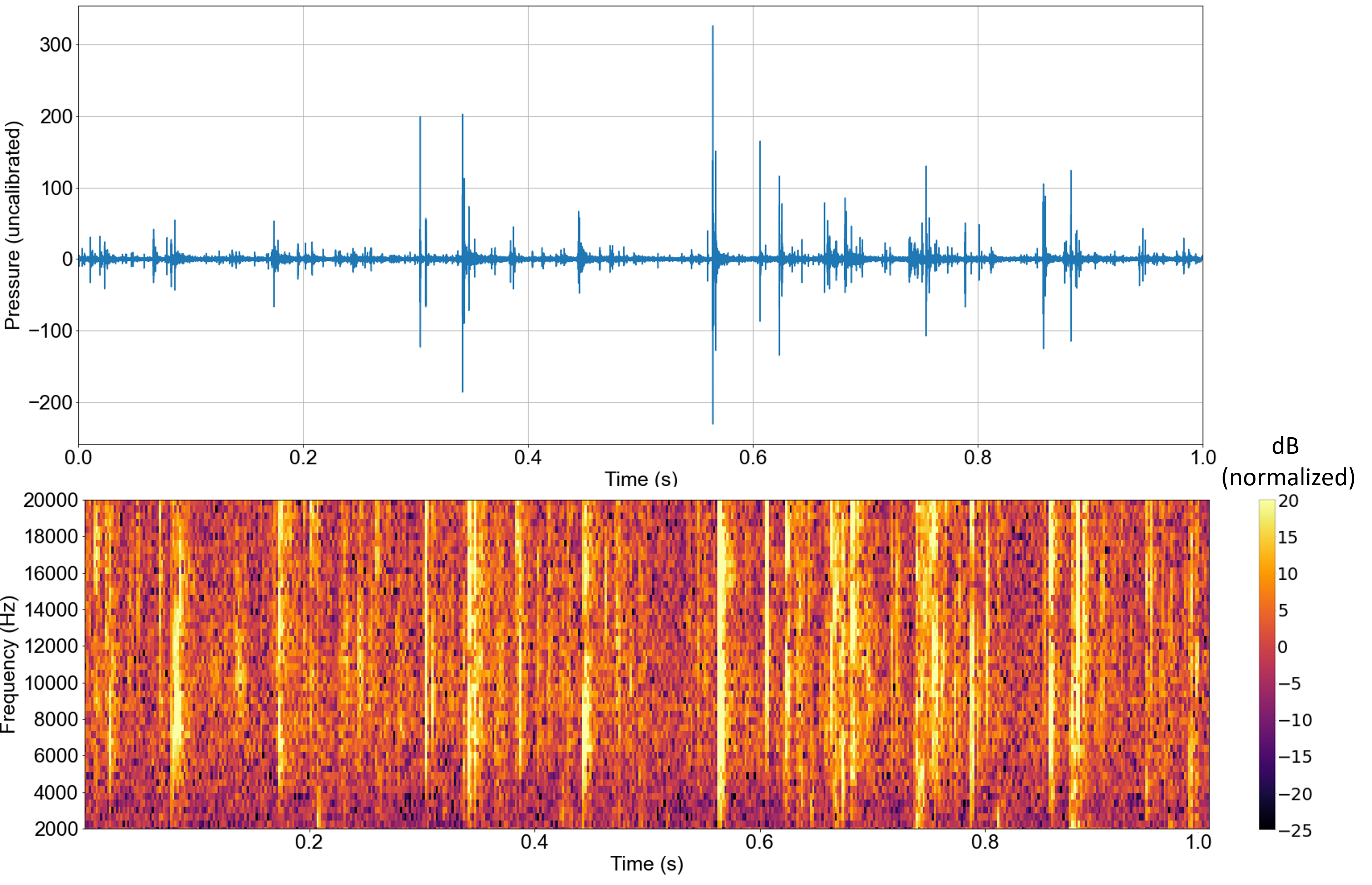}
    \caption{(a) Representative timeseries and (b) spectrogram of snapping shrimp noise background showing broadband impulsive nature of the snaps.}
    \label{fig:snappingshrimp}
\end{figure*}

\section{Challenges and existing methods}\label{sec:challenges}
The acoustic structure of clicks, buzzes and burst-pulses is fundamentally different from that of whistles. It is more naturally described in terms of short-duration waveform events and temporal organization.

Echolocation clicks are short-duration broadband ultrasonic pulses with extremely high peak amplitudes and very short temporal extent, typically on the order of tens of microseconds~\cite{fang2015}. Similar characteristics have also been observed in related delphinid species, where clicks exhibit high source levels and strongly impulsive behavior~\cite{wahlberg2011}. As illustrated in Fig.~\ref{fig:target_signal_examples}(a), click signals are characterized by isolated high-amplitude transients arranged in temporally structured click trains. Their discriminative information is therefore concentrated in local waveform peaks rather than in globally averaged signal statistics.

Burst-pulses and buzzes, likewise, are broadband and impulsive, but differ from clicks in their temporal organization, typically consisting of rapid pulse sequences with smaller inter-pulse intervals~\cite{Wang2022}, and are used more commonly during social tasks~\cite{Wang2022, Sims2012}. Compared with click trains, burst-pulses generally exhibit more complex temporal organization. As shown in Fig.~\ref{fig:target_signal_examples}(b), burst-pulses retain their transient and broadband character, but appear as denser pulse-group patterns. Despite these differences, these signal types share broadband transient characteristics.

Detecting these vocalizations in real-world underwater recordings presents several interconnected challenges:
\begin{enumerate}
    \item \textbf{Train--deploy mismatch:} High-quality positive examples are relatively scarce, but are necessary to provide reliable click/buzz/burst-pulse structure. Usually, the examples available are obtained in controlled environments with known species, recorded with high signal-to-noise ratio (SNR) and noise backgrounds different from ocean noise. By contrast, the target deployment environment consists of low-SNR underwater recordings, where dolphin vocalizations are rare and embedded in complex background noise.
    \item \textbf{Acoustic confusability:} The acoustic environment in Singapore waters is dominated by snapping shrimp, which produce broadband impulsive sounds that can closely resemble dolphin clicks, buzzes and burst-pulses, as shown in the sample 1-second timeseries and spectrogram in Fig.~\ref{fig:snappingshrimp}. So the target and background share the very transient structure a detector must rely on, making frequency-localized discrimination insufficient, as noted in other works as well~\cite{Siddagangaiah2020, Caruso2020}. A recent review identifies snapping shrimp noise and ship cavitation noise as broadband impulsive sources that can be confused with delphinid clicks, underscoring that this confusability is a general obstacle rather and not Singapore- or Southeast Asia-specific~\cite{Gracic2025}.
    \item \textbf{Sensitivity to data handling:} Because discriminative information sits in short-duration waveform peaks rather than in globally averaged statistics, the task is unusually sensitive to normalization, clipping and mixing choices. If these design choices interact poorly they can either remove valuable information needed to discriminate signals from noise, or manufacture an artificial energy gap between classes, letting the model exploit an energy shortcut instead of learning acoustic structure.
\end{enumerate}

\subsection{Prior work}
A recent review highlights click detection methods in whales and dolphins~\cite{Gracic2025}. We refer the reader to this for detailed survey, and only present a subset of related work here. Early detection methods for odontecete vocalizations relied on signal-processing heuristics based on physical characteristics of the signals. For example, Klinck and Mellinger~\cite{Klinck2011} and Madhusudhana et al.~\cite{Madhusudhana2015} applied the Teager--Kaiser Energy Operator (TKEO) to detect echolocation clicks by exploiting their impulsive energy signatures. While effective in relatively clean environments, such methods are sensitive to impulsive interference and lack robustness in snapping-shrimp-dominated conditions. Siddagangaiah et al.~\cite{Siddagangaiah2020} demonstrated that entropy measures can achieve accurate whistle and click detection under controlled noise conditions. However, their study also showed that click detection degrades significantly in the presence of snapping shrimp noise, due to the similarity in impulsive structure. Time-frequency features combined with a support vector machine have been used to classify broadband impulsive signals~\cite{Caruso2020}. However, such approaches rely on predefined features, which limits their ability to capture complex or previously unseen signal patterns.

Deep learning methods, particularly CNNs, are the most common approach for automated detection of marine mammal vocalizations. Most existing systems operate on spectrogram representations. Bermant et al.~\cite{Bermant2019} demonstrated near-perfect detection of sperm whale clicks using spectrogram-based CNNs, while Shiu et al.~\cite{Shiu2020} showed that neural networks can improve recall for whale call detection. In related work, unsupervised learning was used to classify clicks from the Gulf of Mexico~\cite{Frasier2017}, and parameters such as frequency spectra and inter-click-interval (ICI) were used to distinguish the clicks of IPHD versus Indo-Pacific finless porpoises in \cite{Kimura22}, and belugas versus narwhals in \cite{Jones2022}. Bio-acoustic foundation models such as Surfperch~\cite{Williams2025} are also being increasingly used, by fine-tuning with additional classifier head layers to be tailored to specific bio-acoustic applications~\cite{Dumoulin2025, Burns2025}.

Within the taxonomy of methods surveyed in \cite{Gracic2025}, spanning feature-analysis, frequency-content, energy-based, template-matching, adaptive, and supervised/unsupervised learning approaches, TRIDENT falls into the supervised, waveform-domain deep-learning category. Notably, the energy- and feature-based families are more vulnerable to impulsive interference. Hence, the energy- and entropy-based baselines we compare against later are representative of whole classes of methods rather than isolated cases.

The existing literature reveals a gap in the detection of dolphin clicks/buzzes/burst-pulses, especially in shallow tropical waters. Here, the broadband transient structure of the target signals overlaps strongly with that of snapping shrimp noise, making simple energy- or feature-based or frequency-localized discrimination insufficient. The one prior detector developed for Singapore waters \cite{Vishnu2024} targets FM signals and thus, by design, recovered fewer buzzes and burst-pulses. Broadband transient detection in this snapping-shrimp regime thus remains unaddressed.

This work addresses these limitations through a data-centric, waveform-based detection framework. The proposed approach combines controlled SNR-based data generation, waveform-specific normalization, hard negative samples, and training-time augmentation to better deal with the limited training data, the transient nature of vocalizations and the statistical properties of the target noise environment, and address the challenges presented earlier. Furthermore, the testing metrics are tuned towards rare-event detection in long recordings. In this way, the work aims not only to improve detection performance under real-world conditions, but also to reduce reliance on artificial statistical cues that do not generalize to deployment recordings.

\section{Experimental setup and data collection}\label{sec:expdesign}
We deployed single-hydrophone stand-alone LS1 recorders acquired from Loggerhead Instruments \cite{LS1, Vishnu2024, Vishnu2026} across ten island sites (shown in Fig.~\ref{fig:Ls1sites}) for several months intermittently between June 2019 and September 2021 with the aim of studying both reef and marine mammal acoustics. Each recorder contains a High-Tech Inc \cite{HTI} hydrophone with a sensitivity of $-179.90$~dB~re~V/$\mu$Pa, gain of 16.77~dB, sampling frequency of 96~kHz and a bit-depth of 16 bits. Each recorder was deployed at the ocean bottom for approximately 1 month per deployment. In total, we obtained 54.2 recorder-months of data over this period from the island sites, details of which are shown in Table~\ref{tab:datarecorded} (column 4).

\begin{table}
    \centering
    \begin{tabular}{|c||c|c|c|c}
        \hline Location &No. of &No. of&Months\\
         &broadband signals&days with&of data\\
         &detected&detections & recorded \\
        \hline \hline Hantu & 170 & 33 & 7.7\\
        Jong & 13 & 7 & 5.2 \\
        Kusu & 14 & 3 & 4.3\\
        Raffles Lighthouse  & 61 & 5 & 5.2\\
        Seringat & 6 & 1 & 6.3\\
        Semakau Northwest & 0 & 0 & 5.2\\
        Semakau Southwest & 0 & 0 & 2.3\\
        Subar-Darat & 51 & 3 & 6.1 \\
        Subar-Laut & 10 & 1 & 3.9\\
        TPT  & 0 & 0 & 5.6\\
        \hline
    \end{tabular}
    \caption{Island-sites at which acoustic data was recorded, with number of broadband transient vocalizations detected and number of days on which these were detected.}
    \label{tab:datarecorded}
\end{table}


\section{Detector design and training}\label{sec:detector}
The detector development follows a multi-stage pipeline spanning preprocessing, dataset construction, and waveform-based classification. Because the targets are short-duration, high-amplitude transients whose discriminative information lives in local waveform structure, preserving that structure while maintaining numerical stability is a design requirement, and it drives every choice below. We organize the design around four coordinated components: (1) a 1D CNN operating directly on the waveform, avoiding the resolution trade-offs of a time-frequency representation; (2) controlled SNR-based mixing to approximate low-SNR deployment conditions; (3) waveform-specific normalization that preserves transient peaks without manufacturing class separability; (4) hard negative samples for robustness against realistic false positives. 

\subsection{Preprocessing}
In the preprocessing stage, all raw audio recordings are resampled to 44.1~kHz and passed through a high-pass filter with a cutoff of 1~kHz. This suppresses low-frequency background components and emphasizes the high-frequency transient content associated with dolphin vocalizations. The resampling frequency was determined based on two considerations: (1) Most of the training data was available at this sampling frequency. (2) Though dolphin broadband transients may contain energy up to frequencies of around 170~kHz~\cite{Vishnu2022}, in most of the sample signals we observed in the open ocean, most of the high-frequency energy had dissipated and the detectable acoustic energy in these signals was concentrated up to a frequency of 30~kHz, with the rest being under the shrimp noise floor. This is understandable since higher frequencies get absorbed faster, and this effect is especially noticeable at larger ranges. This is the scenario that is harder to tackle, since closer-range signals are likely to be of high-enough SNR to be detectable.

Crucially, and in contrast to the whistle detector of \cite{Vishnu2024}, we deliberately avoid any nonlinear pre-filtering such as clipping anywhere in the pipeline. By forgoing nonlinear denoising we accept a harder raw detection problem than the one faced in FM-whistle detection: the shrimp snaps remain fully present in the input, overlapping the targets in both bandwidth and impulsiveness. We make this trade deliberately. While nonlinear denoising suppresses impulsive shrimp noise and aids FM detection on spectrograms, it also flattens exactly the transient peaks that distinguish dolphin clicks, buzzes and burst-pulses, and would therefore remove the information this detector depends on. Preserving waveform fidelity end-to-end is, in this regime, more valuable than raising the apparent SNR. Instead, we shift the burden of discrimination onto the model itself, which our ablations show is where it can be handled effectively.

\subsection{Detector architecture}
For frequency-localized sounds such as whistles, spectrogram-based 2D CNNs are often a natural choice because the relevant structure is strongly expressed in the time--frequency plane. In the present task, however, the target signals are broadband transient events whose discriminative information is concentrated in short-duration waveform structure rather than narrowband frequency trajectories. In such cases, converting the signal into a spectrogram introduces a time--frequency resolution trade-off that depends on parameters such as window length, FFT size, and overlap. This can blur the fine temporal details that are important for distinguishing dolphin transients from impulsive background noise.

An alternative is to apply 1D CNNs directly to raw waveforms. This avoids the need for explicit time--frequency parameter selection and allows the model to learn features directly from the waveform itself. For broadband impulsive signals, relatively little additional discriminative information may be gained by projecting the signal into a spectrogram representation, whereas preserving the original time-domain structure can better retain the transient peaks and pulse repetition patterns and organization that characterize clicks, buzzes and burst-pulses. This is especially relevant in snapping-shrimp-dominated environments, where both targets and background noise contain broadband impulsive content and are therefore not easily separable by frequency-concentrated cues alone.

Prior work has shown that sufficiently deep 1D CNNs can achieve performance comparable to spectrogram-based models while offering improved flexibility for raw waveform modeling~\cite{dai2017, kim2026}. These observations motivate the use of a waveform-based 1D CNN in the present work, where preserving transient temporal structure is more important than exploiting a hand-designed time--frequency representation.

\begin{figure}[htbp]
    \centering
    \includegraphics[width=0.8\linewidth]{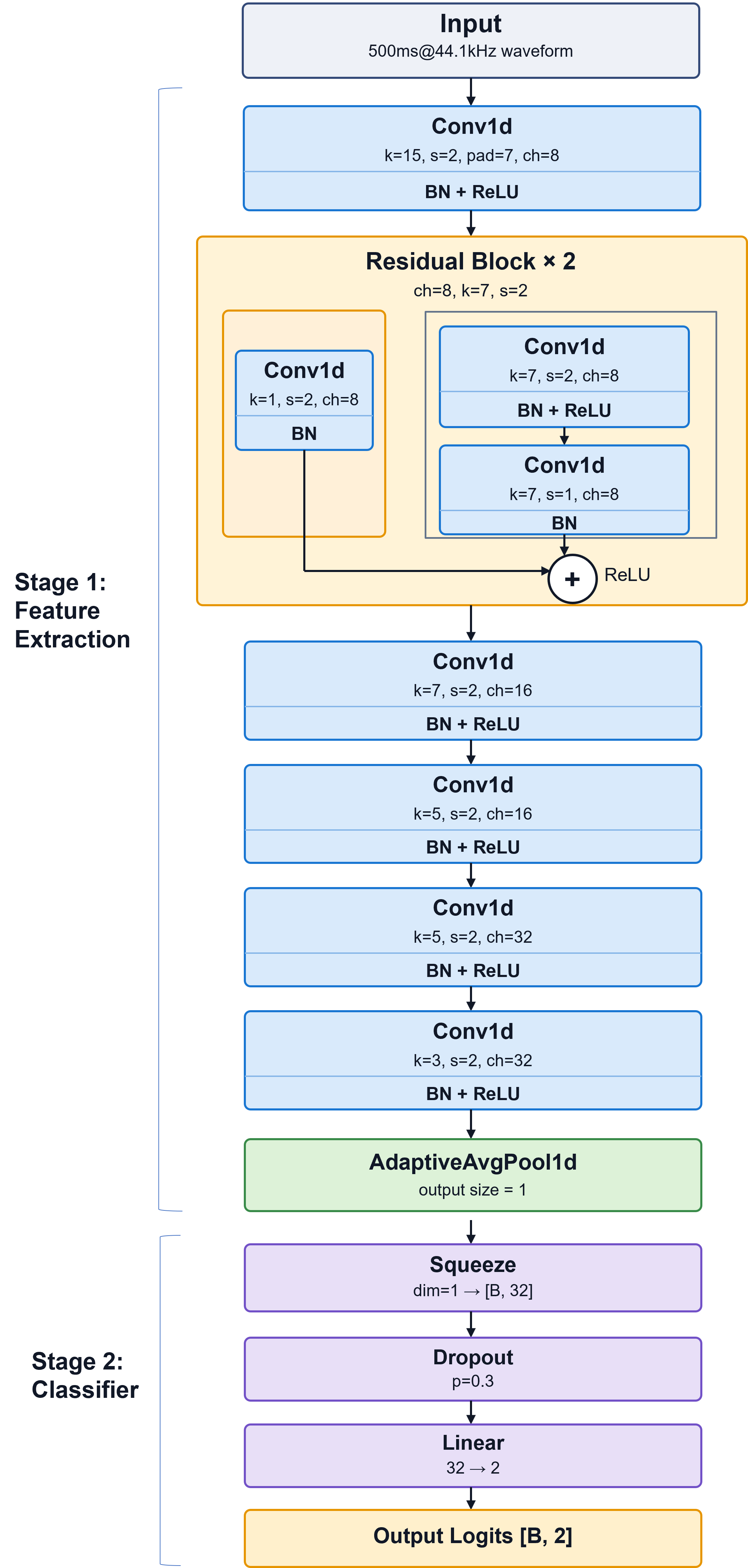}
    \caption{Architecture of the proposed 1D CNN detector TRIDENT. The network operates directly on a 500~ms raw waveform inputs and uses a deep but narrow design with one residual block, progressive temporal downsampling, adaptive average pooling, and a lightweight classification head. Symbols: $k$ is the kernel size, $s$ is the stride length, $ch$ is the number of filters, B is batch-size}
    \label{fig:model_architecture}
\end{figure}

Our detector is implemented as a lightweight 1D CNN operating directly on raw waveform inputs, with the architecture described in Fig.~\ref{fig:model_architecture}. The input to the model consists of fixed-length 500~ms raw waveform segments, corresponding to 22050 samples each. The model is designed to capture temporal patterns of transient signals while maintaining a balance between representational capacity and computational efficiency. The choice of a 500~ms input window is primarily motivated by the temporal characteristics of dolphin buzzes and burst-pulses. In practice, distinguishing these signals from impulsive background noise depends not only on isolated pulses, but also on the temporal organization of repeated transients, including characteristic ICIs. If the input window is too short, it may contain only a few isolated pulses, making it difficult to distinguish a structured click train or burst pulse from incidental noise. Conversely, excessively long windows increase computational cost, complicate label assignment, and may include multiple unrelated events without providing proportionate benefit. The selected size is sufficiently long to preserve the local temporal structure of dolphin transient signals, while remaining short enough to maintain efficient training and inference. It therefore provides a practical balance between temporal context and computational efficiency.

\subsection{Binary detection classes}
The detection task is formulated as a binary classification problem, where the model distinguishes target segments from non-target segments (background noise and other sources). The output layer yields a single detector score value which can be thresholded to yield a detection decision on whether the sound clip contains (1) dolphin broadband vocalizations, i.e. a `signal' (embedded in noise), or (0) only ambient noise and sounds that are not dolphin broadband vocalizations.

All 3 signal types (clicks, buzzes and burst-pulses) are broadband transient events and share certain impulsive structural characteristics. At the same time, they are not identical in their detailed temporal structure, and the effectiveness of this unified formulation must therefore be evaluated experimentally. Whether clicks, buzzes and burst-pulses should be modeled jointly or by signal-specific detectors is itself a design question, which we evaluate experimentally in Section~\ref{sec:results}. In the joint setting, a single model detects all three types within one binary framework; in the signal-specific setting, we train a click-only and a buzz+burst-pulse detector independently. Anticipating the results, the joint formulation performs on par with the specialized detectors while requiring only a single model at inference, and we adopt it as our mainline approach. Its rationale is twofold:
\begin{enumerate}
    \item A unified detector is encouraged to learn the shared broadband-transient structure across vocalization types rather than overfitting to one, while remaining robust to variability within the target class.
    \item It allows us to deploy a single detector to capture these broadband transient vocalizations in long-term recordings, reducing the inference-time effort.
\end{enumerate}

\subsection{Model training and evaluation}
Training is done using PyTorch in Python, and the Adam optimizer is used with a learning rate that decreases with the number of epochs. No pretrained models are used. During training, a softmax is added to the output layer to limit its output within the [0, 1] range, and the cost function used for training is the cross-entropy between the detector's output and a one-hot encoded output vector 
indicating whether a signal was present in the input timeseries or not. We add batch normalization (BN) after each convolutional layer, and dropouts during training to improve generalization.

Finally, we evaluate the trained model is evaluated on held-out datasets using threshold-independent and application-oriented metrics. In particular, we assess performance using the probability of detection/recall at a probability of false alarm of 1\% (PD@PFA=1\%), following the Neyman-Pearson criteria \cite{Kay1998}. The area under the curve (AUC) of the receiver operating characteristics (ROC) is also considered as a secondary metric. While AUC summarizes the performance of the detector over a range of PFA and detection thresholds, PD@PFA=1\% is particularly relevant for long-term monitoring, where even a modest false alarm rate can translate into a large number of false detections over extended recordings, thereby increasing the burden of downstream review and analysis~\cite{Vishnu2024}. Operating at a low PFA helps keep the volume of shortlisted detections at a manageable level for subsequent tasks such as abundance estimation. Thus, selecting a model that works in this regime of operation helps ensure the detector is tuned towards our application.

\subsection{Data sources}
\paragraph{Positive samples} include dolphin transient signals collected from the following sources:
\begin{itemize}
    \item recordings of IPHD from Sentosa pools, done by the Acoustic Research Laboratory~\cite{Seekings2010},
    \item recordings of IPBD from Ocean Park Hong Kong by Acoustic Research Laboratory~\cite{Vishnu2022}, 
    \item recordings of Irrawaddy dolphins and Australian humpback dolphins (\textit{Sousa sahulensis}) from Cornell Lab of Ornithology Macaulay Library~\cite{MacaulayAus, MacaulayIrra}, and
    \item recordings of common bottlenose (\textit{Tursiops truncatus}) obtained from online dataset provided by~\cite{DiNardo2023}, 
    \item online open repositories \cite{Sayigh2016, NOAA2023} containing clips spanning 17 dolphin species and 2 porpoise species.
\end{itemize}
These sources provide a diverse set of cetacean vocalizations recorded under different species, environments, and acquisition conditions, with specific focus on species found in the Southeast Asia region. The high-SNR clips are mostly used to construct the positive data for training and validation sets, while the test set is deliberately constructed to emphasize more challenging conditions using most of the low-SNR clips.

\paragraph{Negative samples} consist of two categories:
\begin{itemize}
    \item background noise segments, obtained from real underwater noise recordings, typically containing shrimp noise but also other sources including ships,
    \item hard negative samples (false positives), identified through false-positive mining and manual verification, typically consisting of non-target signals with buzz-like, impulsive, or high-frequency transient characteristics.
\end{itemize}

A key finding in this line of work is the importance of hard negative samples. Incorporating challenging non-target examples during training substantially improves detection robustness~\cite{Shiu2020}, highlighting the need for realistic training data distributions rather than relying on clean or overly simplified negative classes. Compared with ordinary noise segments, hard negatives are more challenging, often because they share structural similarities with dolphin vocalizations. Their inclusion is intended to improve robustness against false detections in real-world recordings. In total, 159 hard negative segments were used in the final experimental dataset, as summarized in Table~\ref{tab:dataset_split}.

Later, as vocalizations of interest were detected in the initial set of underwater recordings, they were incorporated into the training dataset so that the detector could be retrained and fine-tuned to search for vocalizations from local dolphin species. Each such iteration of the detector involved first manually sorting through the clips scored the highest by the detector from amongst a subset of the newly obtained recordings spread across time and geographical locations, and annotating the true positives and false alarms. The top-scored 100 clips per day were considered for the additional annotation. Annotators also listened to the recordings in the vicinity of true positives to pick up possible missed detections. These additional annotated sound clips were added into the existing training/validation dataset.

Table~\ref{tab:dataset_split} summarizes the composition of the final mixed-target dataset used in the mainline configuration of this work. For the target-definition experiments reported later, the same data sources, preprocessing pipeline, and negative sample construction are retained, while the composition of the positive class is adjusted according to the task definition. The training-validation-test data split was done roughly in a 50:15:35 ratio, while making adjustments to ensure there were enough samples from each subclass of positives in the three categories. The validation data set contains fewer raw samples compared to the test set, because in the validation set, positives are further augmented via synthetic mixing (as described in the next subsection), whereas this is not done in the test dataset.

\begin{table}[htbp]
\centering
\caption{Composition of the final mixed-target dataset used in the mainline configuration of this work. Pos indicates independent click, buzz and burst-pulse samples, which are mixed with noise in different permutations to produce samples for training and validation, and Neg indicates independent negative data samples (non-signals).}
\label{tab:dataset_split}
\begin{tabular}{lrrrrr}
\toprule
Split & Pos & Pos & Neg & Neg & Total \\
& (Click) &(Buzz+Burst-pulse) & (Noise) & (Hard) & independent \\ 
& & & & & samples \\
\midrule
Train      & 723 & 808 & 3390 & 93 & 5014 \\
Validation & 187  & 196  & 700 & 33 & 1116  \\
Test       & 491  & 247  & 1893 & 33 & 2664  \\
\bottomrule
\end{tabular}
\end{table}

\subsection{Training data preparation and augmentation}
The data consists of fixed-length windows generated synthetically by randomly mixing available positive signals with the noise in every iteration using controlled SNR-based mixing. The motivation for this approach is that very few recordings of vocalizations typical of Singapore waters were available beforehand. Hence, the data is compiled by mixing the available vocalization recordings and noise to develop a detector capable of handling a shrimp-dominated noise background. This helps us efficiently use the available data and obtain maximum data diversity for model development. As an indicator of the diversity of data, we use about 3.6~million unique permutations of the signal+noise samples for training and validation. Data augmentation techniques, such as shifting and amplitude scaling, are further applied to improve robustness and generalization. Flipping-based augmentation is not applied here because click, buzz, burst-pulse and shrimp snaps are not symmetric in the time domain, and we want the detector to be able to use this information in detection if possible. 

The SNR of the synthesized data is defined as:
\[
\mathrm{SNR}_\mathrm{peak-to-noise} = 20 \log_{10}\left(\frac{\mathrm{Peak}_{\mathrm{signal}}}{\mathrm{RMS}_{\mathrm{noise}}}\right)
\]
where \(\mathrm{Peak}_{\mathrm{signal}}\) denotes the maximum absolute amplitude of the pre-mix positive signal after peak normalization, and \(\mathrm{RMS}_{\mathrm{noise}}\) denotes the root mean square of the noise segment. This formulation based on peak of the signal is in contrast with the regular RMS signal term usually used in L2-norm based SNR calculations. In RMS-based formulations, strength is measured over the entire waveform window, and hence it remains an appropriate descriptor for background noise energy. For the signal, the peak-based definition is chosen here because the target signals in this work are transient and peak-dominated. For dolphin broadband transients in particular, the discriminative information is concentrated in short-duration high-amplitude peaks, and a peak-based signal term therefore better reflects their perceptual and structural salience than an RMS-based formulation. Their overall energy over the whole window is not of consequence, as short bursts of click/buzz/burst-pulse energy are often detectable even if not spread throughout the entire window. In subsection~\ref{sec:results} we briefly discuss the improvement obtained from this design decision.

However, choosing this signal definition also means we need to use a relatively larger range of values of $\mathrm{SNR}_{\mathrm{peak-to-noise}}$ to reflect realistic challenging scenarios, because the peak of a transient signal is much larger than its RMS value over a larger time window. Thus, this SNR formulation and SNR range cannot be tuned independently. Instead, they must be jointly designed to ensure that the generated data reflects realistic acoustic conditions. Based on this intuition and experiments done with the data, clicks are mixed using an $\mathrm{SNR}_{\mathrm{peak-to-noise}}$ range of \([15.6, 35.6]\)~dB, while burst-pulses and buzzes are mixed using a range of \([20.4, 40.4]\)~dB, which were found to work best for detection performance. These would correspond to smaller SNR values if computed using the typical RMS-based approach.

\subsection{Normalization}
The normalization strategy significantly affects whether the model learns transient structure or a global-energy shortcut. Though a normalization scheme based on the background noise statistics alone may be argued, it leads to the possibility of the detector using energy-based cues for detection, as we show in Section~\ref{sec:results}. We therefore normalize the positive (signal+noise) segments by their median-absolute-deviation \textit{after} mixing, and negative segments are normalized directly by the same metric. This keeps the RMS distributions of the positive and negative classes substantially overlapping (Fig.~\ref{fig:rms_shortcut}), so the detector cannot separate them on energy alone. Since clipping/nonlinear denoising is avoided throughout as described earlier, waveform dynamics are preserved end-to-end.

Amplitude scaling is introduced as a training-time data augmentation strategy to improve robustness to variations in signal magnitude. During training, each input data waveform (after synthesis and normalization) is multiplied by a random scaling factor sampled from a range of \([0.8, 1.2]\), without modifying its SNR. This augmentation is applied dynamically during data loading and is enabled only for training samples, while validation and test data remain unchanged. This simulates amplitude variations that may be caused by factors such as recording gain differences, hydrophone sensitivity variations and acoustic propagation effects. By exposing the model to a range of amplitude scales, this approach encourages the learning of scale-robust features rather than reliance on absolute signal magnitude alone.

The validation and test datasets are non-overlapping with the training set. The test data contains 2900 samples including both positives and negatives. In order to provide as realistic testing performance as possible, the positive samples in the test set are not generated by synthetic mixing, and instead contain mostly low-SNR samples hand-picked from the original set of available data. The validation data contains 5980 samples, including both negatives and positives in roughly equal numbers, with the positives generated by synthetic mixing. It is used to determine the criterion for stopping training to avoid overfitting --- the training is stopped when the recall evaluated on the validation set begins to reduce consistently, or is saturated. 

\section{Results}\label{sec:results}
We first verify that the normalization strategy of Section~\ref{sec:detector} does not leave an exploitable energy shortcut. Next, we compare joint versus signal-specific target definitions. We then examine the effect of introducing hard negatives, and finally benchmark the detector against established methods from the literature.

\subsection{Effect of normalization}
\begin{figure}[htbp]
    \centering
    \includegraphics[width=0.95\linewidth]{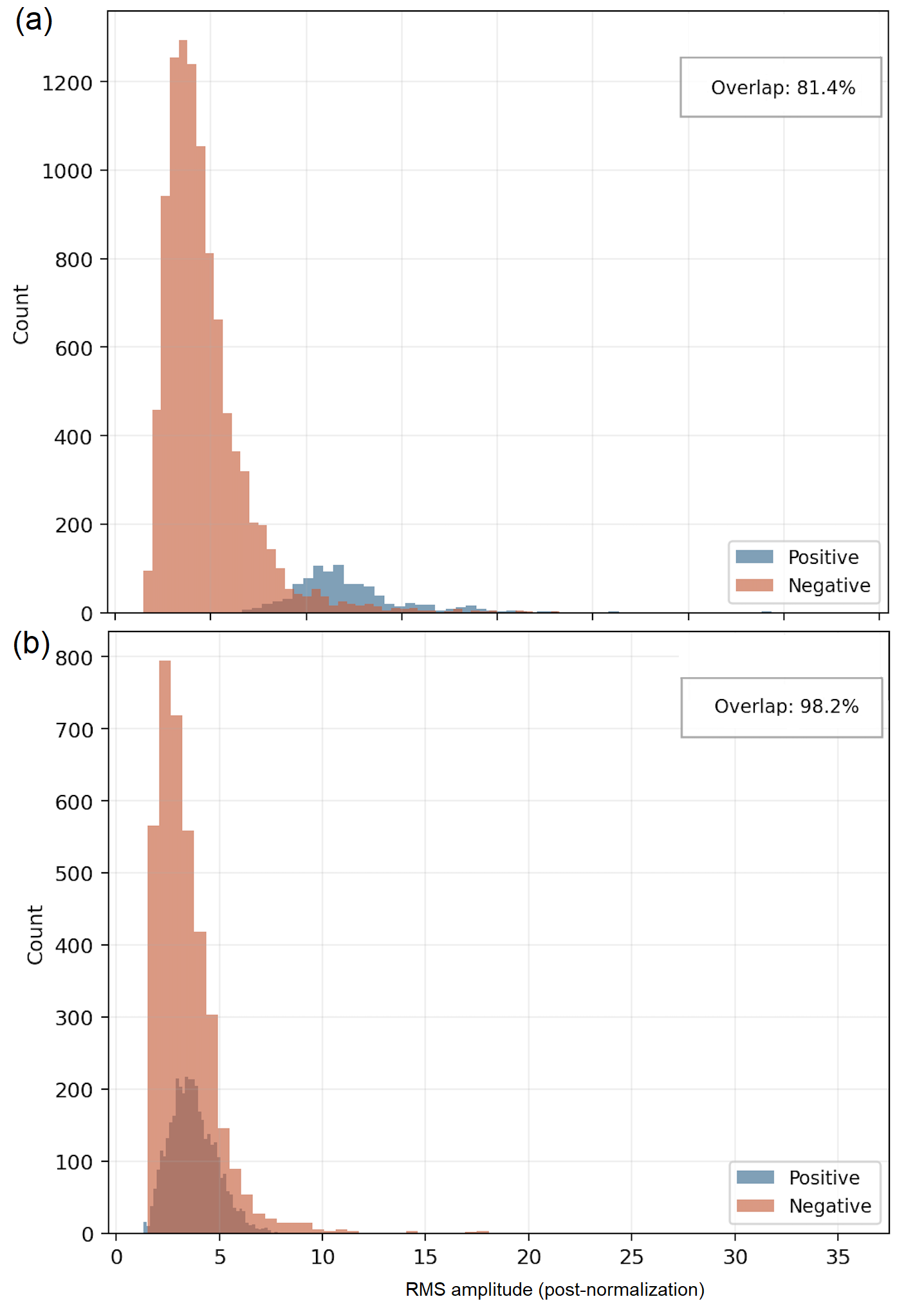}
    \caption{Comparison of RMS distributions of positive and negative sample windows under normalization based on (a) noise background alone (for positives), and (b) proposed normalization based on data. Under the proposed strategy, the overlap between the two distributions increases substantially, encouraging the detector to rely on acoustic structure rather than global amplitude shortcuts.}
    \label{fig:rms_shortcut}
\end{figure}

Fig.~\ref{fig:rms_shortcut} compares the RMS distributions of positive and negative samples under (a) a background noise-based normalization pipeline (which considers normalization based on just the noise segment added to a positive sample), and (b) under the proposed scheme which normalizes after the addition. The earlier pipeline separates the two classes by a pronounced energy gap, which a detector can exploit to reach high apparent performance without learning any acoustic structure. The proposed scheme substantially increases the overlap between the distributions, forcing the detector to rely on transient waveform structure rather than global amplitude. Because this behaviour is standard once normalization is handled correctly, we do not dwell on it further, but it is a prerequisite for the ablation studies that follow to be meaningful.

\begin{figure}[htbp]
    \centering
    \includegraphics[width=\linewidth]{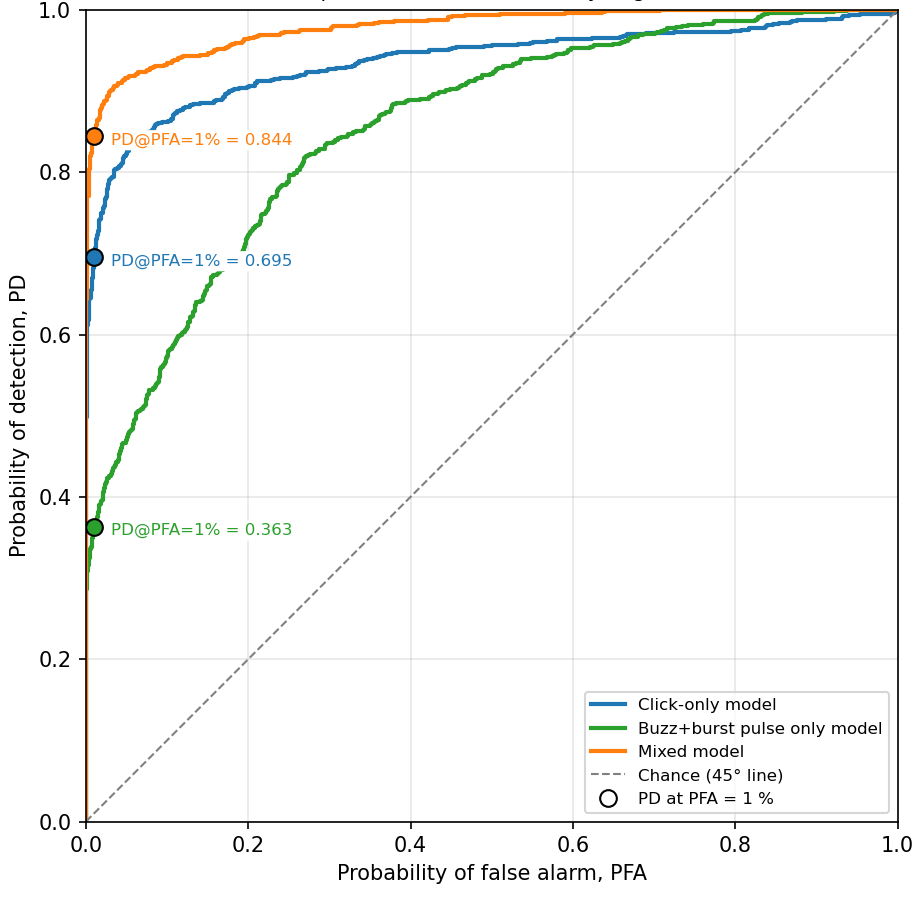}
    \caption{ROC comparison of the click-only, buzz+burst-pulse-only, and joint models on the final mixed-test set. The annotated points indicate PD@PFA=1\% for each model. The joint model achieves the best overall mixed-target performance.}
    \label{fig:roc_target_definition_mixed}
\end{figure}

\begin{figure*}[htbp]
    \centering
    \includegraphics[width=\linewidth]{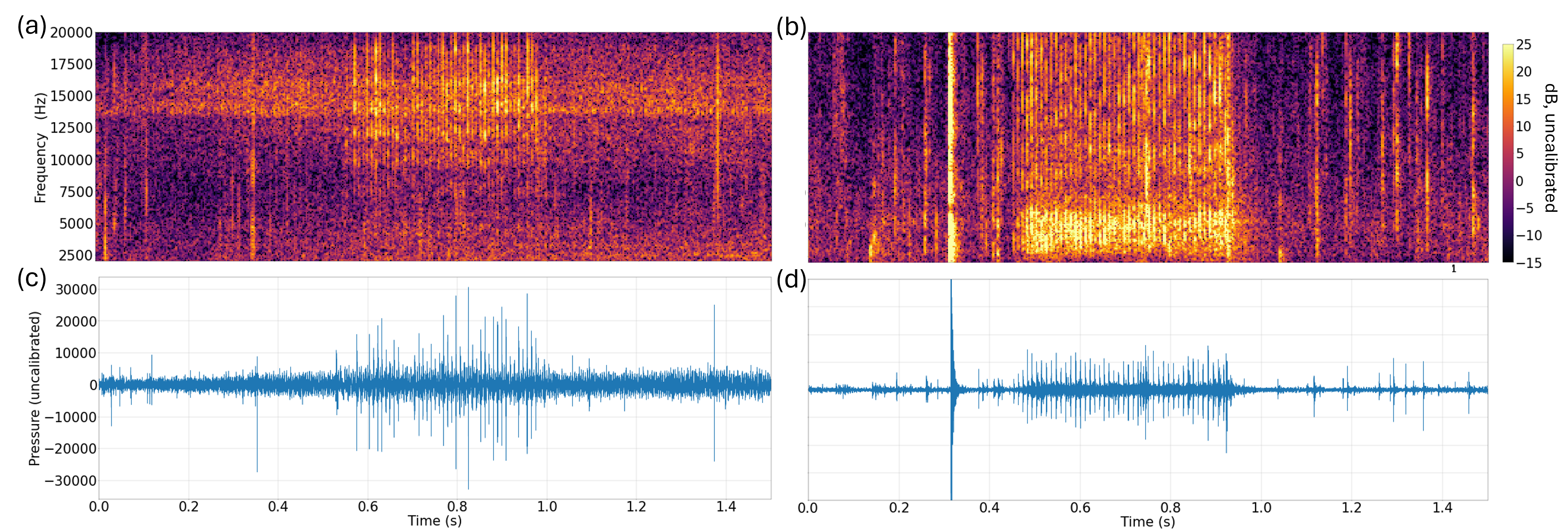}
    \caption{Two examples of hard negatives introduced into the training dataset with features similar to buzzes or squawks, (a) and (b) spectrograms of hard negative clips containing fast boat sounds, and (c) and (d) their corresponding timeseries waveforms showing regular broadband impulsive structure.}
    \label{fig:hardnegatives}
\end{figure*}


\begin{table*}[htbp]
\centering
\caption{Comparison of different ML-based detectors and target definitions under click-only, buzz+burst-pulse-only, and final mixed-test settings. The mixed-test results are reported on the final evaluation set consisting of all three signal types, plus the hard negatives. Bold values in a column indicate the largest row in that column}
\label{tab:target_definition_results}
\renewcommand{\arraystretch}{1.1}
\begin{tabular}{lcccc}
\toprule
Model / Data for metric & \makecell{Click-only PD@PFA=1\%} & \makecell{Buzz+burst-pulse PD@PFA=1\%} & \makecell{Mixed-test PD@PFA=1\%} & \makecell{Mixed-test AUC} \\
\midrule
Click-only                  & 0.786         & 0.514         & 0.695         & 0.936 \\
Buzz+burst-pulse            & 0.092         & 0.903         & 0.363         & 0.854 \\
Click + buzz + burst-pulse  & \textbf{0.811} & \textbf{0.911} & \textbf{0.844} & 0.959 \\
Click + buzz + burst-pulse (no hard negatives)  & 0.752 & 0.870 & 0.791 & \textbf{0.974} \\
Surfperch + Logistic regression & - & - & 0.753 & 0.966 \\
Surfperch + MLP & - & - & 0.782 & 0.969 \\
\bottomrule
\end{tabular}
\end{table*}

\subsection{Impact of target definition and detector design}
We now consider two target signal classes: clicks, which contain transients separated by longer ICIs, and buzz+burst-pulses which contain closer-spaced trains of pulses~\cite{Wang2022}. We study the effect of considering the classes separately in isolation, and together as a single class. Table~\ref{tab:target_definition_results} summarizes the performance of the click-only, buzz+burst-pulse-only, and click+buzz+burst-pulse configurations under matched and mixed evaluation settings. Unless otherwise stated, the mixed-test results reported here correspond to the final click+buzz+burst-pulse evaluation set, which provides a more reliable assessment of mixed-target performance.

Under matched evaluation conditions, the signal-specific models perform moderately well on their respective target types. The click-only model achieves a PD@PFA=1\% of 0.786 on the click-only test set, while the buzz+burst-pulse-only model achieves 0.903 on the buzz+burst-pulse test set. These indicate that specialized models can learn effective representations when the target class is narrowly defined and the evaluation setting is aligned with that definition.

However, this specialization does not translate into robust mixed-target performance. On the final mixed-test setting, the click-only model degrades to a PD@PFA=1\% of 0.695, indicating limited generalization to burst-pulse signals. The buzz+burst-pulse-only model performs even worse, with a PD@PFA=1\% of 0.363, showing that a detector specialized for these patterns alone does not transfer effectively to a mixed-signal detection task.

By contrast, the joint click+buzz+burst-pulse model achieves the most balanced overall performance. Its click-only PD@PFA=1\% (0.811) and buzz+burst-pulse PD@PFA=1\% (0.911) are both better than that of the respective specific models, showing how it has generalized better using information across the different classes. More importantly, it performs best on the final mixed-test setting, achieving the highest PD@PFA=1\% of 0.844 and an AUC of 0.959. This indicates that joint training leads to a broader and more robust representation of dolphin transient vocalizations under realistic mixed-target conditions. This is also reflected in the mixed-test ROC comparison shown in Fig.~\ref{fig:roc_target_definition_mixed}, where the joint model achieves the most favorable overall operating characteristics and the best PD@PFA=1\% among the three configurations. Since the joint detector outperforms the specialist detectors, yet needs only one model at inference, we adopt it for TRIDENT and use it in all subsequent analysis.

Furthermore, our ablation studies reveal that the signal peak based SNR formulation performs 3.5\% better than a model trained using a signal RMS-based formulation (0.815). This elucidates that the SNR definition tuned to the signal characteristics works best for the model's overall performance.

\subsection{Impact of hard negative samples}
Hard negative samples substantially change the detector's behaviour. Unlike ordinary background segments, they represent acoustically challenging non-target patterns that the model genuinely tends to confuse with dolphin vocalizations --- in our recordings, mainly boat- and ship-generated sounds whose broadband impulsive structure resembles click-trains or buzzes (Fig.~\ref{fig:hardnegatives}). Their inclusion forces the model to learn a sharper boundary against these realistic false-positive modes.

Comparing the hard-negatives-only configuration against the baseline in Table~\ref{tab:target_definition_results}, we see that the detector that uses hard negatives in its training obtained about 4\% higher PD as compared to the detector that does not include the hard negatives in its training, showing the clear benefit from inclusion of the targeted training data.

\subsection{Real-time performance and comparison with existing methods}
The computation time to run TRIDENT was benchmarked on a Jetson Xavier NX edge-computing module, by testing on 5 minutes of audio recording. Including the preprocessing and model inference steps, total time taken for inference was, on an average, 8.4 ms for every 0.5~s window of data. This shows that the model's lightweight design renders it feasible for real-time application.

We benchmarked TRIDENT against two families of established approaches. The first comprises generic signal-processing detectors: the entropy-based method of Siddagangaiah et al.~\cite{Siddagangaiah2020} and the TKEO-based method~\cite{Klinck2011}. The second comprises two ML-based networks trained on embeddings from Surfperch~\cite{Williams2025}, a standard foundation model for underwater bioacoustic recordings. The Surfperch model, built on an EfficientNet architecture, was trained on a combination of tropical reef, bird, and other sounds to achieve strong transfer learning on underwater tasks. It is typically used to generate embeddings from audio windows. The Surfperch-based models are trained on the click+buzz+burst-pulse data with hard negatives included, by freezing the embedding layers and adding either (1) a logistic-regression layer, or (2) a multi-layer perceptron (MLP) with two fully-connected layers containing 1280 and 256 neurons. The first layer has ReLU activations and the second has a sigmoid activation, allowing it to be trained using the same metric as TRIDENT. Both are trained using the Adam optimizer. 
Note that the Surfperch-based models are used here only for comparison and benchmarking against the more lightweight TRIDENT
. These models are far bulkier, containing 24.2-24.5 million parameters and $\sim$4500 times larger than TRIDENT, which has only 5362 parameters.

On the mixed dataset of Table~\ref{tab:target_definition_results}, the TKEO-based detector achieved an AUC of 0.30 (PD@PFA=1\% of 0.0005) and the entropy-based detector an AUC of 0.56 (PD@PFA=1\% of 0.0015). Both baselines key on impulsive energy, which in Singapore waters is overwhelmingly shrimp-generated. Their poor performance is therefore not a tuning artifact, but a direct consequence of the target-noise confusability that motivates this work. As the original authors note, neither method was designed for snapping-shrimp conditions, and the gap underscores the need for a detector purpose-built for noisy tropical waters.

The Surfperch-based models form a stronger benchmark
. Table~\ref{tab:target_definition_results} shows that TRIDENT still attains the highest PD@PFA=1\% (0.844), exceeding both the Surfperch+logistic-regression (0.753) and Surfperch+MLP (0.782) models. In terms of AUC, TRIDENT is only marginally below the Surfperch based models. Since PD@PFA=1\% is the operating regime most relevant to long-term monitoring, where a low false-alarm rate governs the downstream review burden, this advantage is the more consequential one for our application. Crucially, TRIDENT achieves it with roughly three orders of magnitude fewer parameters than the Surfperch-based models, likely due to the effectiveness of its architecture and data pipeline tailored specifically to broadband transient dolphin vocalizations rather than relying on a general-purpose embedding.


\section{Discussion and Conclusion}\label{sec:conclusion}
TRIDENT was applied to the long-term recordings available, and detected totally 325 vocalization events (0.5~s windows containing vocalizations) within the duration June 2019-September 2021. The distribution of these vocalizations across the island sites are shown in Table~\ref{tab:datarecorded}, column 2. Notably, no vocalizations have (yet) been detected at Semakau or TPT, whereas there is relatively heavier vocalization activity at Hantu, Sisters Islands, Raffles Lighthouse, and moderate activity at Jong and Kusu. Reassuringly, these acoustic hotspots fall within the same island groups where overlapping occurrence hotspots of IPHD and IPBD have recently been reported by combining scientific and grey literature and citizen science datasets~\cite{Ng2026}, providing independent corroboration of our detections and reinforcing the case for sustained monitoring of these habitats in hyper-urbanized waters.

\subsection{Temporal structure of detections}
Beyond raw detection counts, the \textit{pattern} of detections in time carries information. Across sites, the broadband vocalizations do not arrive uniformly but in tight clusters --- many detections falling within a span of minutes, separated by long silent stretches. Because a single acoustic encounter can thus generate many correlated detections, the raw number of detected signals is a fragile measure of animal presence. We therefore report the number of \textit{days with detections} (Table~\ref{tab:datarecorded}, column 3) as a more robust indicator of visitation, an approach we recommend for comparable long-term PAM studies in this regime.

The detections also show a consistent diel structure, with almost all broadband vocalizations occurring between approximately 07:00 and 20:00. While a fuller spatio-temporal and diel/seasonal analysis is beyond the scope of this methods-focused paper, this daytime concentration is consistent with active foraging or social behaviour and warrants dedicated study. The tightly clustered, predominantly daytime activity reported here only hints at the underlying structure of these detections; a dedicated analysis of their spatio-temporal, cyclical (diel, lunar, seasonal) and behavioural trends is forthcoming, and TRIDENT provides the necessary foundation for it.

\subsection{Generalization across species and sites}
Our training data are deliberately weighted toward the species that most frequently visit Southeast Asia waters, namely IPHD and IPBD (spotted more often in Singapore waters), and Irrawady dolphins. TRIDENT is therefore expected to recover these most accurately and robustly. At the same time, the positive set spans many dolphin and porpoise species, so the detector may well generalize to broadband transients from other cetaceans. This broader capability has not yet been tested and remains an open question. Establishing it would have direct ramifications for cetacean characterization across Southeast Asia, where other species such as finless porpoises have also been documented \cite{Kimura22, Kimura2022a, Ming1990}, and where snapping shrimp noise remains a persistent obstacle to PAM.


\subsection{Practical implications and future work}
TRIDENT's lightweight 1D CNN architecture and its operation directly on the raw waveform make it well suited to on-device inference, as shown by the inference time tests. A natural research trajectory is therefore to embed it within real-time, edge-based PAM systems such as those demonstrated in Singapore waters earlier~\cite{Beng2024}, enabling in-situ detection without storage and post-processing overhead. Such low-complexity, real-time detectors for fixed ocean observatories are also highlighted as a future direction in the broader click-detection literature~\cite{Gracic2025}. At a false-alarm rate of PFA = 1\%, the detector yields a load consistent with feasible quality-control review \cite{Shiu2020} and compatible with such deployments.

Despite these improvements, several limitations, or scope for future work, remain. First, our positive dataset was constructed from a rule-based candidate detector followed by manual verification, which may have introduced selection bias toward more prominent or canonical signal patterns. For example, annotated data for low-SNR scenarios are lacking for use in training, and hence we had to resort to SNR-based mixing for the training instead. Our current model adopts a relatively lightweight 1D CNN architecture, which improves efficiency and training stability but may limit the ability to capture more complex temporal dependencies. Future work could therefore explore more expressive temporal models, such as sequence-based or attention-based architectures, while preserving computational feasibility. These may also capture valuable temporal context, such as the clustering and presence of sounds in relation to other sources (eg. ships), which may improve detection performance. The detector's generalization to cetacean species beyond IPHD and IPBD remains to be tested, with potential value for monitoring across Southeast Asia. Furthermore, its outputs provide the basis for a forthcoming analysis of the spatio-temporal, cyclical and behavioural patterns of dolphin presence in Singapore waters, and lend itself to more applied downstream tasks. This study can plug large gaps in understanding of the dwindling numbers of reported occurrences of delphinid species in these waters~\cite{Ng2026}, and their behavioural patterns with respect to the hyper-urbanized waters of Singapore.

\subsection{Conclusion}
We have presented TRIDENT, a waveform-based 1D CNN detector for dolphin click trains, buzzes and burst-pulses in the snapping-shrimp-dominated waters around Southeast Asia, where the target signals and the dominant noise share broadband impulsive structure and are therefore difficult to separate by frequency-localized or energy-based means. We find that the data pipeline is as important as the architecture: controlled SNR-based mixing, normalization, hard negative mining, and amplitude-scaling augmentation together steer the model toward learning transient temporal structure rather than energy shortcuts. Deliberately avoiding the nonlinear denoising that proved effective for whistles is necessary to preserve that structure - this accepts a harder raw problem in exchange for waveform fidelity. A single joint detector handles all three signal types with negligible loss relative to specialized models. It substantially outperforms generic signal-processing baselines, and, despite being roughly three orders of magnitude smaller, achieves the best detection probability at a 1\% false-alarm rate among all compared methods, including models built on a general-purpose bioacoustic foundation embedding. Applied across 54.2 recorder-months, TRIDENT reveals tightly clustered, predominantly daytime vocalization activity, motivating days-with-detections as a robust presence metric.

\section*{Acknowledgement}
We thank Matthias Hoffmann-Kuhnt and Paul Seekings for acquiring the dolphin vocalization data that was used in training, and Too Yuen Min in obtaining the long-term recordings. We also acknowledge the receipt of media from the Cornell Lab of Ornithology Macaulay Library that was used for training, and thank them for the same. From Macaulay Library, We used specimens ML112562-ML112564 for \textit{Orcaella brevirostris}~\cite{MacaulayIrra}, and ML123171-ML123198 for \textit{Sousa sahulensis}~\cite{MacaulayAus}.

\bibliographystyle{IEEEtran}
\bibliography{megafauna, reef_shrimp}

\end{document}